\documentclass[prl,twocolumn,superscriptaddress,showpacs,preprintnumbers,onlyamsmath,amssymb]{revtex4}

\usepackage{graphicx}

\usepackage{bm}

\newcommand{\laNMR}{$^{139}\rm La$}

\newcommand{\eu}{$\rm Eu$}

\newcommand{\lfsco}{\mbox{$\rm La_{1.65}Eu_{0.2}Sr_{0.15}CuO_4$}}

\newcommand{\lndsco}{\mbox{ $\rm La_{1.48}Nd_{0.4}Sr_{0.12}CuO_4$}}

\newcommand{\lffesco}{$\rm La_{2-x-y}Eu_{y}Sr_xCuO_4$}

\newcommand{\lscoEu}{$\rm La_{1.85-y}Eu_{y}Sr_{0.15}CuO_4$}

\newcommand{\lscoNd}{\mbox{$\rm La_{2-x-y}Nd_{y}Sr_{x}CuO_4$}}

\newcommand{\octa}{$\rm CuO_6$}

\newcommand{\sr}{$\rm Sr$}

\newcommand{\tc}{$T_{\rm c}$}

\newcommand{\tlt}{$T_{\rm LT}$}

\newcommand{\lbco}{\mbox{$\rm La_{2-x}Ba_xCuO_4$}}

\newcommand{\cuo}{$\rm CuO_{2}$}

\newcommand{\islm}{$^{139}T_{1}^{-1}$}

\newcommand{\po}{$\rm P_{0}$}

\newcommand{\Tomega}{$T_{\rm \omega_{0}}$}

\newcommand{\Tc}{$T_{\rm c}$}

\newcommand{\Tg}{$T_{\rm g}$}

\newcommand{\roab}{$\rho_{\rm ab}$}

\newcommand{\roup}{$\rho_{\rm Up}$}

\begin{document}

 \title{Interplay between Freezing and Superconductivity  in the Optimally Doped  \mbox{\lfsco} under Hydrostatic Pressure}

 \author{B.Simovi\v c}

 \altaffiliation{Present address: Laboratorium f\"{u}r Festk\"{o}rperphysik
HPF, ETH H\"{o}nggerberg CH-8093 Z\"{u}rich }
\affiliation{Condensed Matter and Thermal Physics, Los Alamos
National Laboratory, Los Alamos, NM 87545}

 \author{M. Nicklas}

\altaffiliation{Present address: Max Planck Institute for Chemical
Physics of Solids, Noethnitzer Str. 40, 01187 Dresden, Germany.}
\affiliation{Condensed Matter and Thermal Physics, Los Alamos
National Laboratory, Los Alamos, NM 87545}
 \author{P.C. Hammel}
 \affiliation{Department of Physics, The Ohio State University, Columbus, OH 43210 }

\author{M. H\"{u}cker}
\affiliation{Physics Department Brookhaven National Laboratory,
Upton, New York 11973}

 \author{B. B\"uchner}

 \affiliation{RWTH Aachen, Aachen Germany}

 \author{J. D. Thompson}

 \affiliation{Condensed Matter and Thermal Physics,
  Los Alamos National Laboratory, Los Alamos, NM 87545}

 \date{\today}

 \begin{abstract}

We study the electronic properties of a \lfsco\ single crystal
under hydrostatic pressure up to 2.9 GPa. Both the freezing of the
Cu 3d moments and the structural transition from the orthorhombic
(LTO) to the tetragonal (LTT) phase are observed via the
relaxation of the nuclear magnetization of \laNMR\ nuclei.
Resistivity and magnetic susceptibility measurements have been
carried out under pressure on the same sample. The combination of
all data reveals  the connection between glassy dynamics, charge
localization and the disappearance of superconductivity in the LTT
phase.
\end{abstract}



\maketitle

The phenomenon of charge-stripe ordering in hole-doped
antiferromagnets has received much attention since its discovery
in \lndsco\ by elastic neutron
scattering.~\cite{TranquadaJM:Eviscs} Much of this interest has
been motivated by the proximity between stripe order and high
temperature superconductivity and the potential relationship
between these two states.~\cite{Dagotto:NPS2003, Kivelson:HTD2003,
Castro-neto:CIS2003} It is now accepted that competing electronic
interactions can stabilize charge-stripe order, but the question
of whether the  order observed in \lndsco\ is a genuine electronic
phenomenon or specific to a particular distortion of the \cuo\
plane remains controversial.

Evidence for stripe ordering have been reported so far in the
lanthanum cuprates \lscoNd~\cite{TranquadaJM:Eviscs,
Tranquada97:PRL} and more recently in \lbco.~\cite{Fujita02:PRL}
These compounds undergo a first order transition from an
orthorhombic (LTO) to a tetragonal (LTT) phase which affects
superconductivity dramatically at optimal hole
doping.~\cite{AxeJD:Strilc} In both phases, the \octa\ octahedras
tilt collectively toward the c-axis of the perovksite
structure.~\cite{AxeJD:Strilc} This distortion produces a
staggered buckling of the \cuo\ plane with a symmetry different
for LTO and LTT phases.~\cite{AxeJD:Strilc, Simovic:Local} Stripe
order emerges in lieu of superconductivity when the tilt exceeds
$\approx 3.6 ^\circ$ in the LTT phase.~\cite{TranquadaJM:Eviscs,
BuchnerB:Cribds}

Despite those observations, the reason for the disappearance of
superconductivity is not understood. Neutron-scattering
experiments in \lscoNd\ have shown that the stripe arrangement
mimics, to some extent, the symmetry of the tilt pattern of the
LTT phase, and the detection of the stripe order in this compound
is coincident with the first order structural
transition.\cite{TranquadaJM:Eviscs} This suggests strong pinning
of the charge-stripe order by the underlying lattice, which tends
to localize charge carriers. Yet, this picture appears incomplete
for \lfsco\ where signs of ordering of the Cu 3d moments are
observed only below 30K, while the LTT phase occurs at
\tlt=135K.~\cite{KlaussH-H:antosm, Simovic, Simovic:Local} Neutron
scattering has failed so far to find evidence of charge-stripe
order in \lfsco\, but the slow dynamics of the Cu 3d moments,
probed by NMR  at low temperature, has been shown to be consistent
with the picture of a glass-forming charge-stripe liquid
state~\cite{Simovic}. This is indicative of strong frustration,
which could also be detrimental to superconductivity.

To clarify the relation between freezing, superconductivity and
structural distortion, we use pressure to change the buckling of
the \cuo\ plane without modifying doping or substitutional
disorder. We have investigated a \lfsco\ single crystal under
pressure up to 2.9 GPa by means of the magnetization relaxation
rate \islm\ of the \laNMR\ nuclear spins, the magnetic ac
susceptibility $\chi_{\rm ac}$ and the in-plane resistivity \roab.
The pressure-temperature phase diagram of \lfsco\ is obtained from
the pressure dependence of four distinct quantities. First, the
freezing temperature, \Tomega, is defined from the maximum of
\islm where the frequency scale $\Omega$ of the magnetic hyperfine
field fluctuations at the La site equals the Larmor frequency
($\omega_{0} = 50 \rm MHz$) of the \laNMR\ nuclear spins. Second,
the temperature \tlt\ also is determined from \islm through its
response to fluctuations of the surrounding electric field
gradient near the first order structural
transition~\cite{SuhBJ:Locmsp}. Third, the superconducting
transition temperature \Tc\ is determined from the temperature
dependence of $\chi_{\rm ac}$. Fourth, the amplitude of the upturn
in \roab\ observed below 100K  measures the tendency for charge
carriers to localize on lowering temperature. \newline This study
shows as a whole that the disappearance of superconductivity in
\lfsco\ originates from long-time dynamical properties of the
charge inhomogeneous phase. This anomalous long-time behavior
appears to be controlled by the amplitude of the buckling of the
\cuo\ layer which decreases under pressure.  These findings  show
that frustration plays a central role in the sharp suppression of
superconductivity in the LTT phase of \lfsco.

\begin{figure}[tb]
\vspace{0cm}
\begin{center}
\includegraphics[width=0.9\linewidth]{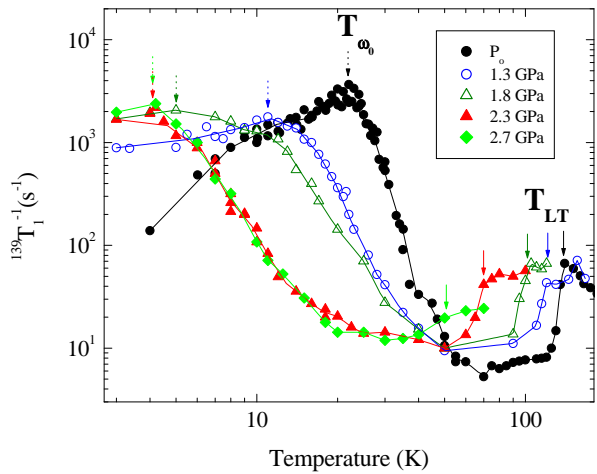}
\caption{\islm versus temperature measured at different pressure.
\tlt\ and \Tomega\ are marked with solid and dotted arrows
respectively. We find that $dT_{\omega_0}/dP = -9 \ \rm
K.GPa^{-1}$ up to $P = 1.8 \rm GPa$. The same fitting procedure as
proposed in ref~\cite{Simovic} has been used to extract the
relaxation rates from the magnetization recovery}
\end{center}
\end{figure}

This study has been carried out on the same single crystal used in
ref~\cite{Simovic,Simovic:Local}. Two similar clamp-type cells
were used to generate hydrostatic pressure and, in both cases,
fluorinert served as pressure medium. The pressure inside the cell
was determined at room temperature by measuring the resistivity of
a calibrated manganin gauge located next to the sample. Lowering
the temperature from 300 to 4K induces pressure losses of $\approx
0.3 \rm GPa$.~\cite{Thompson:cell84}
\newline The $^{139}$La($I = 7/2$) NMR measurements were  made on
the central \mbox{($m_I = +\frac{1}{2}
\leftrightarrow-\frac{1}{2})$} transition.\islm\ was measured up
to 2.7 GPa by monitoring the recovery of the magnetization after
an inversion pulse in a field of  83.5 kOe. After completing NMR
experiments, the resistivity and magnetic susceptibility
measurements were carried out simultaneously up to 2.9 GPa.

Fig.1 shows the temperature dependence  of the relaxation rate
\islm\ for different values of pressure. The features observed at
ambient pressure (\po) have been examined already and discussed in
detail~\cite{Simovic}. They can be summarized as follows. The
discontinuous drop of \islm\ around \tlt $\approx 135$ K indicates
the occurrence of the first order structural transition from LTO
to LTT. The broad peak in the relaxation rate \islm\ manifests a
gradual freezing of the hyperfine field induced at the La site by
the Cu 3d localized moments. We thus read from Fig.1  the pressure
dependence of \tlt\ and \Tomega\ and we observe that both decrease
continuously with increasing pressure. It is striking, however,
that \Tomega\ saturates near 3K for $P\gtrsim$ 1.8 GPa while \tlt\
continues to decrease at higher pressure. Fig.1 makes it obvious
that \islm(T) below 30 K is unchanged between 2.3 and 2.7 GPa,
whereas \tlt\ is reduced by a factor of two. These differences
establish the remarkable fact that the properties of the freezing
of the Cu 3d moments no longer depend on the underlying lattice at
high pressure.

\begin{figure}[tb]
\vspace{0cm}
\begin{center}
\includegraphics[width=0.95\linewidth]{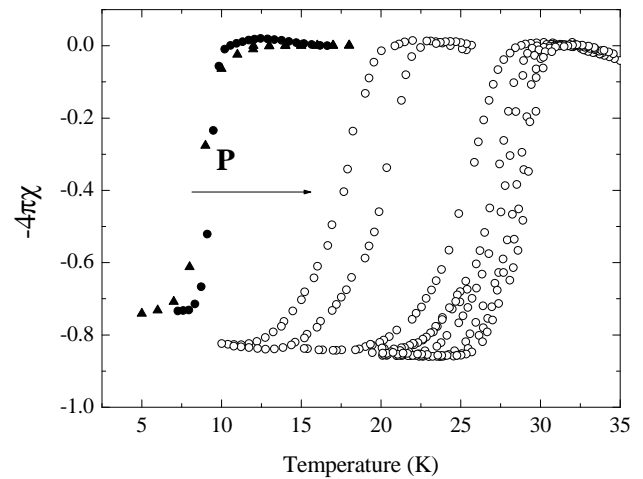}
\caption{Diamagnetic shielding versus temperature for the
successive following pressure (empty circles): 0.67, 1.17, 1.66,
1.75, 1.97, 2.27, 2.51 and 2.9 GPa. At \po, the diamagnetic
shielding has been measured after depressurization from 3.0 GPa
(full circles) and the measurement at \po\ was repeated  4 months
later (full triangles).}
\end{center}
\end{figure}

When looking at Fig.1, a question arises as to whether the
temperature of the glass transition, \Tg, remains finite or drops
to zero under pressure. It should be emphasized that \Tg\ is
conceptually different from \Tomega. Insofar as \Tomega\ indicates
only the temperature at which $\Omega \approx \omega_0$, it can
relate to dynamical properties at thermal equilibrium whereas \Tg\
marks the onset of a fully static non-equilibrium state, the
system being frozen out into a glass. \newline The relation
between \Tomega\ and \Tg\ is {\em a priori} unknown but the two
temperatures clearly differ at \po\ where \Tg$\lesssim \rm 5K$ and
\Tomega$\approx 25\rm K$~\cite{Simovic}. From \islm\ alone, we can
not determine the value of \Tg\ since \islm\ measures only the
spectral weight of the fluctuations at frequencies close to
$\omega_{0}$. Nevertheless, it can be seen in Fig.1 that the broad
maximum of \islm\ occurs around 3K at high pressure. This  means
that $\Omega \approx \omega_{0}$ in this temperature range while
the system is frozen out at \po. Therefore, \Tg\ has dropped
significantly and, hence disappearance of the glass phase under
pressure is conceivable but remains to be proven. \newline This
said, one should keep in mind that despite the pressure dependence
of \islm\ revealing a considerable enhancement of the Cu 3d
moments fluctuations with increasing pressure, dynamics still
remains extremely slow ($\approx 10^{-7}\rm eV$) compared to
typical electron time scales. In particular, at low temperature,
dynamics is certainly faster at pressures higher than \po\ but
would nonetheless appear as static on the time scale ($\gtrsim$1
meV) of neutron-diffraction experiments. Thus, the variation seen
in Fig.1 reflects properties of the inhomogeneous phase {\em over
long time scales} i.e $\tau \gtrsim \omega_{0}^{-1}$.

\begin{figure}[tb]
 \vspace{0cm}
\begin{center}
\includegraphics[width=0.8\linewidth]{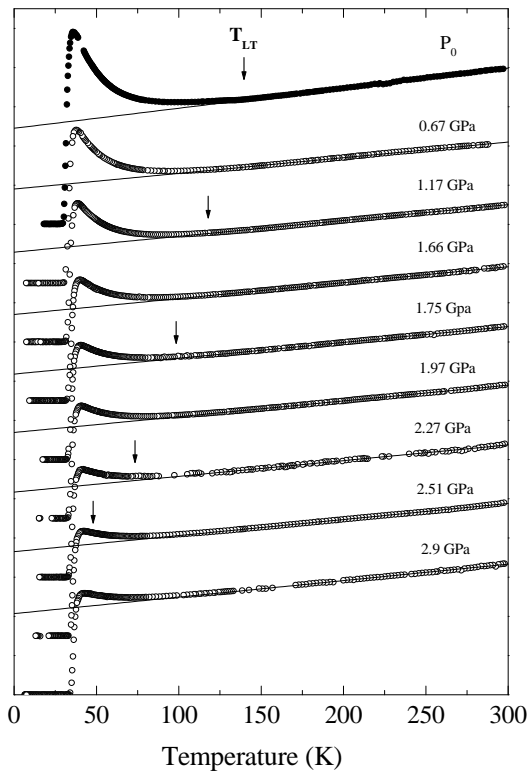}
\caption{$\rho_{ab}$ versus temperature at \po (full circles) and
at higher  pressure (empty circles).  $\rho_{ab}$ at \po\ was
measured after depressurization from 3.0 GPa. The continuous lines
are linear fits described in the text and the solid arrow shows
the temperature \tlt\ when available from NMR data. The sharp drop
at 30K shows filamentary superconductivity in our sample.}
\end{center}
\end{figure}

We show the amplitude of the diamagnetic shielding  in Fig.2.
Shielding of the applied magnetic field (equal to 10 Oe) manifests
the presence of superconductivity in our \lfsco\ single crystal
and Fig.2 makes clear that \Tc\ increases under pressure. We find
that $dT_{\rm c}/dP = 12 \ \pm \ 1 \ \rm K.GPa^{-1}$ for $P <
P_{s} =1.5 \ \pm \ 0.15 \ \rm GPa$ and reduces to $0.6 \ \pm \ 0.2
\ \rm K.GPa^{-1}$ above. We emphasize, however, that the pressure
dependence of the volume fraction of superconductivity cannot be
inferred unambiguously from the data of Fig.2.

The resistivity \roab\, measured in the \cuo\ plane,  is shown in
Fig.3. We analyze the data  as follows. For each value of
pressure, we fit \roab\ in the temperature interval 150-300K to
the linear form $\rho_{0} + \gamma T $ and extract the quantity
\roup(T)$ =\rho_{\rm ab(T)} -\rho_{0}-\gamma T$. We find that the
residual term $\rho_{0}$ varies smoothly with pressure. At \po,
$\rho_{0}= 2.5 \ \pm \ 0.001 \ \rm m\Omega.cm $ and decreases at
the rate of $-0.13 \ \rm m\Omega.cm.GPa^{-1}$. The slope of the
linear term, $\gamma = 5.08 \ \pm \ 0.02 \ \rm \mu \Omega.cm$ at
\po\ and drops to $4.03 \ \pm \ 0.02$ at 0.67 GPa, which is the
first pressure point in our experiment. $\gamma$ remains unchanged
at higher pressure, the variation being less than $0.4\%$.\newline
This linear term in  the resistivity is typical of the metallic
phase of optimally doped cuprates. It is considered to be an
evidence for a non-Fermi liquid behavior, which might be related
to a quantum-phase transition at $T=0$.~\cite{Valla99}In \lfsco\,
it can be seen in Fig.3 that this feature is unaffected by
pressure.

 Most of the effect of increasing pressure on \roab\ is
reflected in the amplitude of the upturn. This upturn is
characteristic of the LTT phase around optimal \sr\ doping and its
amplitude increases with rare-earth
substitution.~\cite{BuchnerB:Cribds, Hucker:Thesis}. Because
superconductivity is suppressed also with increasing rare-earth
content, the upturn might be related to a competing order, but the
mechanism by which localization occurs is still unknown.
Nevertheless, we can estimate the amplitude of the upturn from
Fig.3 as the maximum of the term \roup. We find that it decays as
$e^{-P/P_{\rm s}}$ with $ P_{\rm s} = 1.5 \ \pm \ 0.1 \ \rm GPa$,
as shown in the lower panel of Fig.4.

\begin{figure}[tb]
 \vspace{0cm}
\begin{center}
\includegraphics[width=0.9\linewidth]{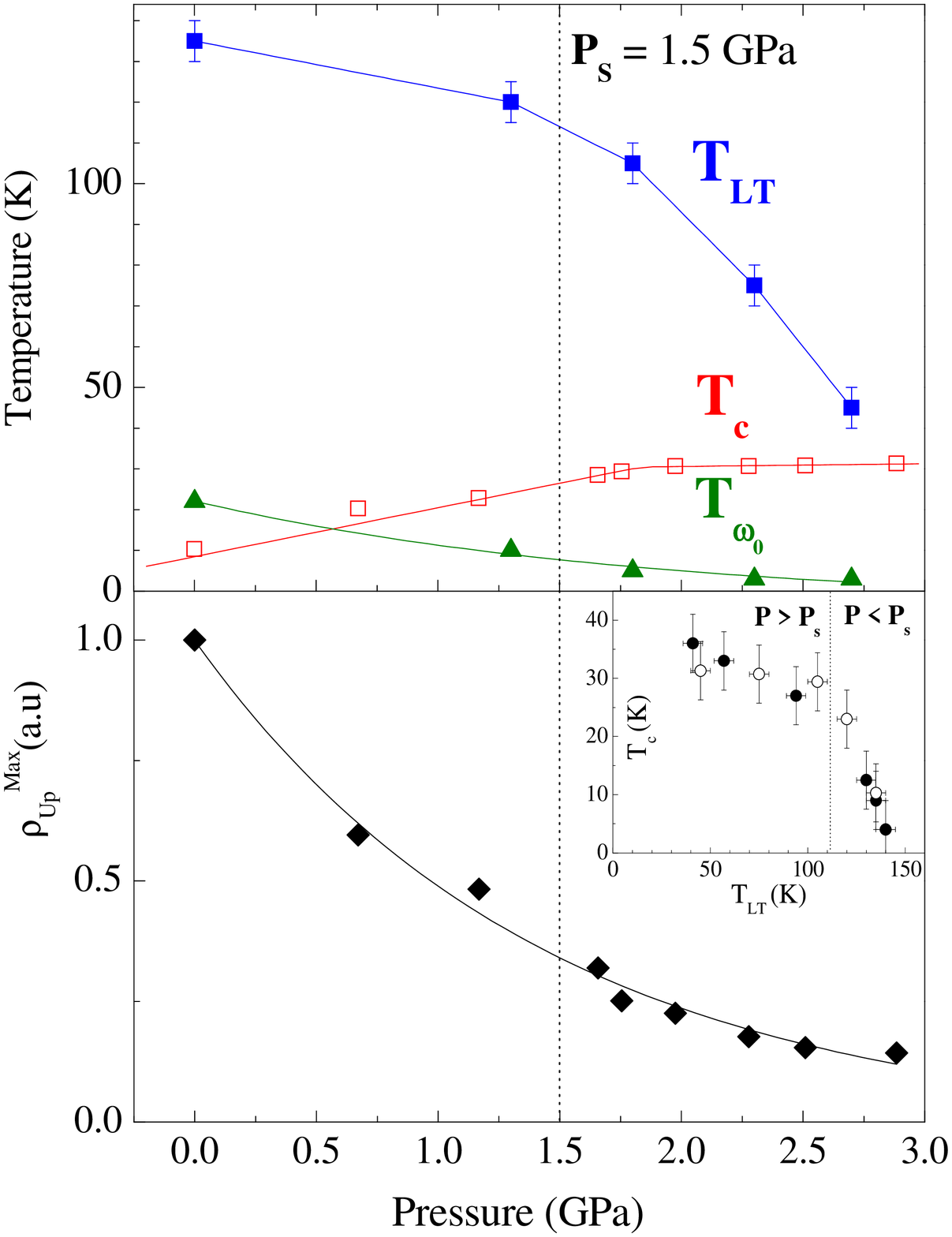}
\caption{Overall phase diagram of \lfsco. Upper panel: pressure
dependence of \tlt\ (full squares), $T_{\rm c}$ (onset) (empty
squares) and $T_{\omega_{0}}$ (full triangles). For $P< P_{\rm
s}$, $dT_{\rm LT}/dP = -11.5 \ \rm K. GPa^{-1}$ and for $P> P_{\rm
s}$, $dT_{\rm LT}/dP = -66 \ \rm K.GPa^{-1}$. Lower panel: maximum
of \roup\ (see text) as a function of pressure. \roup\ has been
normalized to its value at \po\ and the continuous line is a fit
to $e^{-P/P_{\rm s}}$ with $ P_{\rm s} = 1.5 \pm 0.1 \ \rm GPa$
Inset: \tc\ versus \tlt\ measured in \lscoEu\ for different values
of $y$ ranging between 0.03 and 0.3 (full
circles)~\cite{Hucker:Thesis} and versus pressure in \lfsco (open
circles) obtained in the present work. }
\end{center}
\end{figure}

 The relevance of $P_{s}$ becomes clear when comparing  \tlt, \Tc, \Tomega\ and the maximum of \roup\ as a
function of pressure in Fig.4.  We discern two distinct regimes on
either side of $P_{s}$, which differ in the way variations of
\tlt\ alter the electronic properties of the \cuo\ layer. To
better appreciate the role of the lattice, we compare \Tc\ and
\tlt\  as a function of \eu\ content in
\lffesco~\cite{Hucker:Thesis} with that obtained in \lfsco\ under
pressure, in the inset of Fig.4. The correspondence between the
two sets of data is clear: increasing pressure is equivalent to
reducing \eu\ content and, by the same token, to the buckling of
the \cuo\ layer~\cite{BuchnerB:Cribds}. It is, therefore, the
amplitude of the structural distortion that controls the interplay
between freezing and superconductivity in \lfsco. However, the two
compete with each other only when $P \leq P_{\rm s}$, where \tlt\
varies slowly with pressure. Above $P_{\rm s}$, the two phenomena
no longer interfere and, both freezing and superconductivity
depend only weakly on pressure. Since the temperature \tlt\ is
connected to the amplitude of the staggered buckling of the \cuo\
layer~\cite{BuchnerB:Cribds}, the sharp contrast above $P_{\rm s}$
between  the rapid drop of \tlt\ and the saturation of \Tc\ and
\Tomega\, reveals the important conclusion that the coexistence
between slow dynamics of the Cu 3d moments and superconductivity
at high pressure is electronic in origin. Also, the slow dynamics
and the presence of buckling of the \cuo\ layer do not limit \Tc\
which reaches its optimal value above $P_{\rm s}$. It is then
clear that the key insight provided by our work is that the
mechanism of the dramatic suppression of \Tc\ in \lfsco\ is to be
found in the long time ($t \geq \omega_{0}^{-1}$) dynamical
properties of the inhomogeneous phase at low pressure or
equivalently, at high buckling amplitude.

At \po, we have shown that the onset of  slow dynamics of the Cu
3d moments below 30K  can be understood in terms of a
glass-forming stripe-liquid phase.~\cite{Simovic} Following this
idea, a possible interpretation of the decrease of \Tomega\ under
pressure is that frustration  causing charge stripes to freeze out
into a glass is gradually released under pressure. The reduction
of the structural distortion is more likely to increase the
mobility of charge carriers in the \cuo\ layer, thus enhancing the
fluidity of the electronic matrix.~\cite{Dimashko:PRB99} This idea
is also well supported by the fact that most of the reduction of
the resistivity upturn is found where \Tomega\ decreases, which
implies a connection between freezing and localization of the
charge carriers. Insofar as the slow dynamics is driven by slowly
moving charge stripes in the \cuo\ plane, we can infer from the
present data that glassiness, rather than stripes, is what
dominantly competes with superconductivity in  \lfsco. This leaves
open the  question of what makes charge stripes glassy. The
pressure dependence of the glass transition temperature \Tg\,
particularly in relation to the qualitative changes occurring
around $P_{\rm s}$, remains to be investigated to test these
ideas.

To conclude, we report a detailed investigation of the structural
and electronic properties of a \lfsco\ single crystal under
pressure by means of \laNMR-NMR, resistivity and magnetic ac
susceptibility measurements.  Our primary result is clear evidence
that the sharp suppression of superconductivity in \lfsco\
originates from anomalous long-time dynamics of the charge
inhomogeneous phase. This reveals  the key role of frustration in
the interplay between static inhomogeneities and superconductivity
in cuprates.

The work at Los Alamos National Laboratory and Brookhaven was
performed under the auspices of the US Department of  Energy
(Contract No. DE-AC02-98CH10886).

\bibliographystyle{apsrev}

\bibliography{bibtex2}

\end{document}